\documentclass[10pt,conference]{IEEEtran}
\usepackage{todonotes}
\usepackage{amssymb}

\ifCLASSINFOpdf
\else
\fi
\usepackage{algorithm}
\usepackage[noend]{algpseudocode}
\usepackage{algorithmicx}
\ifCLASSOPTIONcompsoc
  \usepackage[caption=false,font=normalsize,labelfont=sf,textfont=sf]{subfig}
\else
  \usepackage[caption=false,font=footnotesize]{subfig}
\fi
\usepackage{url}

\begin{document}
%
\title{Bayesian Verification under Model Uncertainty}

\author{\IEEEauthorblockN{Lenz Belzner}
\IEEEauthorblockA{Institute for Informatics\\
LMU Munich\\
Email: belzner@ifi.lmu.de}
\and
\IEEEauthorblockN{Thomas Gabor}
\IEEEauthorblockA{Institute for Informatics\\
LMU Munich\\
Email: belzner@ifi.lmu.de}
}


%


\maketitle

\begin{abstract}
Machine learning enables systems to build and update domain models based on runtime observations. In this paper, we study statistical model checking and runtime verification for systems with this ability.
Two challenges arise: (1) Models built from limited runtime data yield uncertainty to be dealt with. (2) There is no definition of satisfaction w.r.t. uncertain hypotheses.
We propose such a definition of subjective satisfaction based on recently introduced satisfaction functions.
We also propose the BV algorithm as a Bayesian solution to runtime verification of subjective satisfaction under model uncertainty. BV provides user-definable stochastic bounds for type I and II errors.
We discuss empirical results of a toy experiment. 
\end{abstract}


%
\IEEEpeerreviewmaketitle

\section{Introduction}
Statistical approaches to model checking and runtime verification exploit a domain model in order to evaluate system properties at design and runtime \cite{KNP11}. The system simulates potential traces based on the domain model in order to establish some statistical guarantees about properties of interest.

Statistical verification is often based on a singular domain model \cite{legay2010statistical,jha2009bayesian,zuliani2010bayesian}. Machine learning enables systems to build and adapt models about their application domains based on runtime observations (see e.g. \cite{mackay2003information,Goodfellow-et-al-2016}). In particular, Bayesian statistics generally allow to infer and reason about an infinite amount of models \cite{jaynes2003probability}. Bayesian approaches allow to quantify the likelihood of a particular model, given prior beliefs and observed data. A system that verifies itself at runtime has to cope with model uncertainty to establish reliable verification results: Which hypothesis to assume when assessing system properties?

Model uncertainty induced by learning from limited runtime information also raises another issue: What does it exactly mean for a system to satisfy a particular property, given many model hypotheses and their respective plausibilities?


In this paper, we study statistical verification for systems that are able to build and update their models based on runtime information. The paper's contributions are twofold:
\begin{itemize}
	\item We propose a definition of \textit{subjective satisfaction} for systems that perform runtime verification based on limited information and possibly infinite hypothesis spaces.
	\item We also propose a Bayesian verification algorithm, BV, that enables learning systems to decide on satisfaction or violation of required subjective satisfaction. BV provides user-definable stochastic bounds for type I and II errors (false negatives/positives). By construction, these error bounds are \textit{independent} of the number of observations a system made about its environment.
	\item We empirically establish the validity of BV's error bounds on a toy example.
\end{itemize}

The paper is structured as follows. Section \ref{sec:preliminaries} recaps Bayesian model checking and satisfaction functions for systems with parametrized models. In Section \ref{sec:subjective satisfaction} we introduce our definition of subjective satisfaction. In Section \ref{sec:bv} we discuss the Bayesian treatment of verification under model uncertainty with the BV algorithm. In Section \ref{sec:results}, we describe setup and results of an empirical evaluation of BV. We discuss related work in Section \ref{sec:related}. Section \ref{sec:conclusion} concludes and discusses venues for further research.


\section{Preliminaries}
\label{sec:preliminaries}
This Section recalls Bayesian model checking and satisfaction functions for parametrized models.

\subsection{Bayesian Model Checking}
\label{sec:BMC}

Bayesian model checking (BMC) is based on Bayesian sequential hypothesis testing, and aims to infer the posterior distribution of the probability that a system satisfies its requirements \cite{jha2009bayesian,zuliani2010bayesian}. In contrast to point estimation (e.g. maximum likelihood), the Bayesian posterior captures the uncertainty about the true probability that arises from only performing a finite number of system assessments.

Requirements may be formally specified in a suitable probabilistic temporal logic \cite{pnueli1977temporal,baier2008principles}.

BMC treats a bounded simulation run of a system with a particular configuration as a Bernoulli experiment: The run may either satisfy or violate requirements. As the simulation captures probabilistic domain dynamics, the result of a simulation run is Bernoulli distributed with a probability $p \in [0, 1]$.

BMC infers a posterior distribution over $p$ based on the observed simulation results and a prior assumption about the distribution of $p$. In general, the posterior is proportional to the likelihood of observed data $D$ (i.e. the results of simulation), multiplied by the prior distribution $P(\theta)$ over the parameters of interest, $\theta = p$ in the case of BMC (Equation \ref{eq:bayes}).
\begin{equation}
	\label{eq:bayes}
	P(\theta | D) \propto P(D | \theta) P(\theta)
\end{equation}

BMC models the uncertainty about $p$ by a Beta distribution, the conjugate prior of the Bernoulli distribution. This approach ensures that the posterior is of the same form as the prior distribution, and thus enables efficient sequential updating of the distribution.
The Beta distribution is parametrized by two parameters $a, b \in \mathbb{R}^+$. In the case of BMC, $a$ and $b$ are given by the successes and failures of the simulation runs. Given $s$ successes, $f$ failures, and assuming a uniform prior over $p$, the posterior (for $\theta = p$) is determined by Equation \ref{eq:beta}.
\begin{equation}
\label{eq:beta}
P(p | D) = \mathrm{Beta}(s + 1, f + 1)
\end{equation}

%


Termination can be determined by assessing whether the probability mass above or below the required value of $p_\mathrm{req}$ meets a particular confidence requirement $c_\mathrm{req} \in (0 ; 1)$. For alternative termination criteria, we refer to \cite{zuliani2010bayesian}.

\subsection{Satisfaction Functions}
\label{subsec:sat_function}

Many modern systems operate with models of the environment that are stochastic and parametrized, e.g. models build by machine learning. Classical statistical model checking algorithms, including classical BMC, enable to assess requirement satisfaction for a single parametrization of the model. Recently, the \textit{satisfaction function} was introduced as a concept to allow for efficient, regressive assessment of requirement satisfaction for parametrizable models with potentially infinitely many parameters \cite{bortolussi2016smoothed}. At its core, the satisfaction function is defined as follows.
\begin{equation}
	\label{eq:sat_function}
	f_\mathrm{sat}(\theta) = P(\mathrm{sat} | \theta)
\end{equation}
Here, $\mathrm{sat}$ denotes a boolean variable indicating requirement satisfaction or violation, and $\theta$ are the model parameters. The satisfaction probability is depending on the particular parametrization of the model. However, note that the definition of the satisfaction function does not make any assumptions about the distribution of the parameters themselves. We will now turn to combine estimations about the parameters and the satisfaction function in order to define what we label \textit{subjective satisfaction}.

\section{Subjective Satisfaction}
\label{sec:subjective satisfaction}

Consider a system that was able to make a limited number of observations about the dynamics of its environment. For example, consider a mobile agent whose moves may fail to have an effect with Bernoulli probability $p_\mathrm{fail} \in [0; 1]$. The agent may observe whether its moves are effective or not. Consider a situation where the agent observed its moves 10 times, out of which two had no effect. The following questions naturally arise:

\begin{itemize}
	\item What is $p_\mathrm{fail}$?
	\item How confident can the agent be in its estimate of $p_\mathrm{fail}$?
\end{itemize}

With these two questions in mind, consider now the situation that the agent finds itself in a grid world with obstacles at particular positions. Also, the agent has a sequence of movements to be executed in order to fulfill some given task, e.g. computed by a planning component. Consider that there is a requirement that the agent is only allowed to hit a limited number of obstacles (e.g. 2), with at most a specified probability $p_\mathrm{req} \in [0 ; 1]$.
Another question arises:

\begin{itemize}
	\item What is the probability $p_\mathrm{sat}$ that the sequence of movements will satisfy the requirements, given the limited observations about $p_\mathrm{fail}$?
\end{itemize}

In this setup, an agent has to cope with various uncertainties:

\begin{enumerate}
\item \textit{Domain uncertainty} is inherent to the environment, in our example given by $p_\mathrm{fail}$. It is \textit{aleatoric}, therefore irreducible and originates from the physical setup of the domain (e.g. sensory abstraction, laws of physics, etc.). Note that domain uncertainty in combination with requirements uniquely defines a satisfaction function (cf. Section \ref{subsec:sat_function}).
\item \textit{Model uncertainty} is the \textit{epistemic} uncertainty about the aleatoric domain uncertainty. It arises from the limited number of observations that the agent is able to collect from its environment. Note that model uncertainty not only arises from models learned at runtime. All empirically assessed models convey this kind of uncertainty, in particular all models built with machine learning approaches, regardless of the position in the a system's development lifecycle.
\item \textit{Subjective satisfaction}, the uncertainty about a plan satisfying (or violating) a requirement in a particular situation, is also epistemic. It is a consequence of domain and model uncertainty, and the given system requirements.
\end{enumerate}

The relation of domain and model uncertainty can be modeled in a Bayesian way. This is a widely adopted view, and a vast body of literature and techniques exists for estimating model uncertainty $P(\theta | D)$ based on available domain observations $D$ \cite{mackay2003information,jaynes2003probability}. For readability, we write $P(\theta)$ for $P(\theta | D)$ in the remainder of the paper.

We now combine model uncertainty with the satisfaction function to define subjective satisfaction $p_\mathrm{sat}$.
\begin{equation}
\label{eq:subjective}
	p_\mathrm{sat} = \int_\theta P(\mathrm{sat}| \theta) P(\theta) d\theta
\end{equation}

Subjective satisfaction $p_\mathrm{sat}$ can be interpreted as the parameter of a Bernoulli distribution that models uncertainty about satisfaction of the requirements. Intuitively, Eq. \ref{eq:subjective} weights the satisfaction probability for given parameters w.r.t. the the probability that these parameter represent the ground truth. Subjective satisfaction is considering all possible hypothetic domain parametrizations at once, and weights their respective satisfaction probabilities according to their plausibility (which is based on domain observations).

\section{Bayesian Verification under Model Uncertainty}
\label{sec:bv}

We now define Bayesian Verification (BV), an algorithm for estimating subjective satisfaction by Monte Carlo simulation. By taking a Bayesian stance, we also get a confidence measure for this estimate. In fact, due to assessment of satisfaction with a limited number of simulations, an additional source of uncertainty arises: The uncertainty about the estimate of $p_\mathrm{sat}$. BV establishes and updates a probability distribution $P(\mathrm{sat})$ to quantify this uncertainty, and uses it to decide on termination.
BV takes the following inputs.
\begin{itemize}
	\item The current system state $s$.
	\item $P(\theta)$, the system's model uncertainty.
	\item A probabilistic simulation model of the domain dynamics $M$, parametrized by $\theta$. $M$ takes a state, a plan, a requirement and a parametrization, and yields a boolean variable indicating requirement satisfaction. I.e. this model implicitly provides the satisfaction function $P(\mathrm{sat} | \theta)$.
	\item The system's plan $\pi$ to be assessed.
	\item A system requirement $\phi$, e.g. a temporal logic formula.
	\item A required probability $p_\mathrm{req}$ of satisfying $\phi$.
	\item A required confidence $c_\mathrm{req}$ in the estimate of $p_\mathrm{sat}$.
\end{itemize}

BV is shown in Algorithm \ref{alg:BV}. BV first initializes it estimate of $p_\mathrm{sat}$. As satisfaction of a requirement in a stochastic domain can be interpreted as a Bernoulli random variable we use a uniform prior, which is a $\mathrm{Beta}(1, 1)$ distribution (line 2).

We define the confidence in the estimate of $p_\mathrm{sat}$ that is above the required satisfaction probability $p_\mathrm{req}$ by determining the probability mass of $P(p_\mathrm{sat})$ above $p_\mathrm{req}$. 
\begin{equation}
\label{eq:c_sat}
	c_\mathrm{sat} = \int\limits_{p_\mathrm{req}}^{1} P(p_\mathrm{sat}) d p_\mathrm{sat}
\end{equation}
BV updates its estimate $P(p_\mathrm{sat})$ and uses it in order to decide whether the estimate of satisfaction (or violation) can be done with at least required confidence (cf. Equation \ref{eq:c_sat}). To this end, it performs the following steps in repetition.
\begin{enumerate}
	\item A sample parametrization is drawn from the model uncertainty $P(\theta)$ (line 4).
	\item A simulation run is performed w.r.t. state, plan, requirement and parameters (line 5). Note that the simulation result is distributed accounting for both model uncertainty and satisfaction function, as the parameterization has been sampled from model uncertainty before. That is, $\mathrm{sat} \sim P(\mathrm{sat}| \theta)P(\theta)$ at this point.
	\item The simulation result is used to update the belief distribution about $p_\mathrm{sat}$ (line 6).
	\item The probability mass of the belief distribution is used to determine whether satisfaction or violation have been assessed with at least required confidence (Eq. \ref{eq:c_sat}). If so, the algorithm terminates accordingly (lines 7 and 8).
\end{enumerate}

\begin{algorithm}[H]
	\begin{algorithmic}[1]
		\Procedure {BV}{$s, P(\theta), M, \pi, \phi, p_\mathrm{req}, c_\mathrm{req}$}
		\State $P(p_\mathrm{sat}) \gets \mathrm{Beta}(1, 1)$
		\Loop
		\State $\theta \sim P(\theta)$
		\State $\mathrm{sat} \sim M(s, \pi, \phi, \theta)$
		\Comment $\mathrm{sat} \sim P(\mathrm{sat}| \theta)P(\theta)$
		\State update $P(\mathrm{sat})$ according to $\mathrm{sat}$
		\Comment Eq. \ref{eq:beta}
		\If {$c_\mathrm{sat} \geq c_\mathrm{req}$}
		\Return true
		\EndIf
		\If {$1 - c_\mathrm{sat} \geq c_\mathrm{req}$}
		\Return false
		\EndIf
		\EndLoop
		\EndProcedure
	\end{algorithmic}
	\caption{The \textsc{BV} algorithm for Bayesian verification under model uncertainty.}
	\label{alg:BV}
\end{algorithm}

%

\section{Empirical Results}
\label{sec:results}

We empirically assessed BV on a toy example. While we modeled a very simple example, it may be worth noting that in general the Bayesian approach to model uncertainty scales up to much larger models. There exist varied and powerful tools for sampling from complex, high-dimensional posteriors $P(\theta)$, such as Markov Chain Monte Carlo (see e.g. \cite{diaconis2009markov} for a very interesting read), or variational inference (e.g. \cite{wainwright2008graphical}).

\subsection{Setup}

The state $s$ is constituted by a 10 x 10 grid world, with the agent at position (0, 0). Obstacles are randomly positioned, at an obstacle to free position ratio of 0.2. The agent is presented a plan $\pi$ (an action sequence) of 10 movements (up, down, left, right, with obvious semantics).
The agent has a Bernoulli action failure probability $p_\mathrm{fail}$ uniformly sampled from $[0 ; 1]$. Action failure results in the inverse movement (e.g. failing up yields down). The agent is presented a number of observations about its failure probability before running BV. We build model uncertainty $P(\theta)$ about $\theta = p_\mathrm{fail}$ with a Beta distribution (cf. Eq. \ref{eq:beta} and Section \ref{sec:bv}).

In our setting, $\phi$ is the requirement to hit less than three obstacles while executing the plan. We set $p_\mathrm{req} = 0.9$. This means we allow the agent to classify a plan as satisfying the requirement if it hits less than three obstacles in ninety percent of executions. We use a confidence requirement of $c_\mathrm{req} = 0.95$.

We approximate the ground truth satisfaction probability of a plan $\pi$ by taking the maximum likelihood estimate of satisfaction probability based on 10000 simulation runs. We assessed two error types.
\begin{itemize}
	\item A type I error is an incorrect rejection. This occurs if a plan $\pi$ satisfies $\phi$ with at least probability $p_\mathrm{req}$ and is falsely rejected.
	\item A type II error is a false accept. This occurs if a plan $\pi$ violates the requirements (i.e. $\phi$ is not satisfied with at least probability $p_\mathrm{req}$) and is falsely accepted.
\end{itemize}

We also assessed a variant of BV that does not explicitly build model uncertainty from observations, but rather builds a corresponding maximum likelihood estimate ($\hat{p}_\mathrm{fail} =$ observed failures / number of observations). Line 4 is correspondingly changed to $\theta \gets \hat{p}_\mathrm{fail}$ in Algorithm \ref{alg:BV}.

Our implementation of the setup and BV is available at \url{https://github.com/jazzbob/bv}.

\subsection{Results}

\begin{figure}
	\includegraphics[width=0.49\columnwidth]{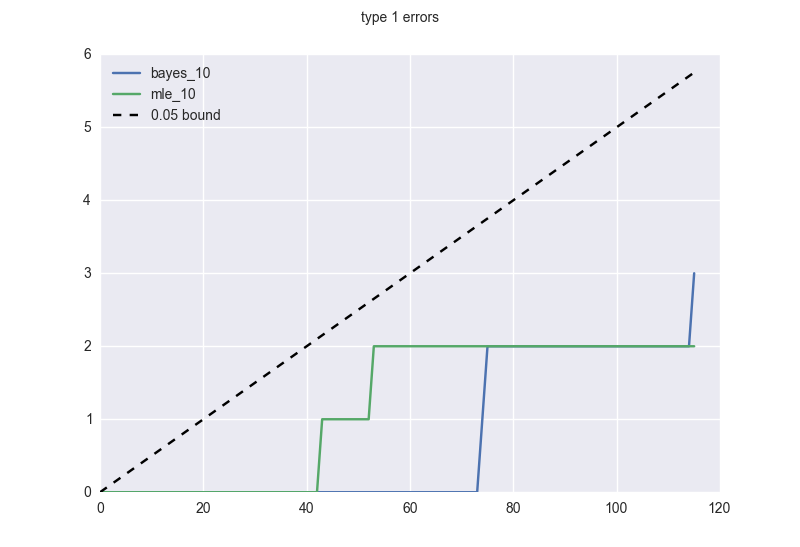}
	\includegraphics[width=0.49\columnwidth]{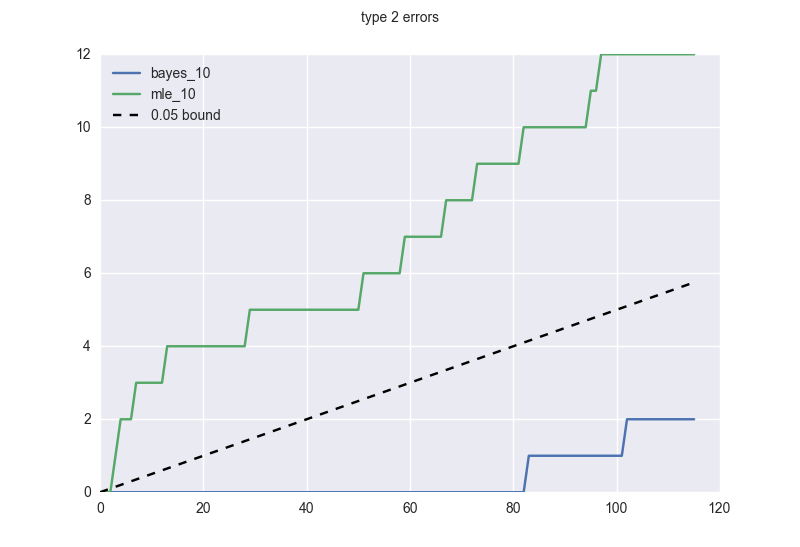}
	\caption{Type I (left) and II (right) errors. X-axis shows number of tested situations ($s$, $\pi$). Vertical axis shows accumulated number of type I and II errors.}
	\label{fig:errors}
\end{figure}

Results are recorded for 10 randomly sampled observations of action failure probability. An exemplary result of our experiments is shown in Figure \ref{fig:errors}. The former shows accumulated type I errors over the course of different setups (i.e. randomly generated environments paired with random plans), the latter type II errors respectively. The dashed line shows the required statistical error bound (0.05 for $c_\mathrm{req} = 0.95$).

In particular, BV is able to establish the required statistical error bounds for both error types, while the MLE approach that is not explicitly using model uncertainty for inference fails to do so for type II errors. We observed this behavior for various numbers of observations presented to the system.


%
%
%


\section{Related Work}
\label{sec:related}

BV is an instance of statistical model checking in general \cite{legay2010statistical}, and Bayesian statistical model checking in particular \cite{jha2009bayesian,zuliani2010bayesian}. Typically, these approaches are assuming a perfect available model, and do not deal with explicitly quantified epistemic model uncertainty.
One of the starting points of the current article is the work on smoothed model checking \cite{bortolussi2016smoothed}. SMC approximates a satisfaction function w.r.t. uncertain model parameters by Gaussian process regression. However, SMC does not incorporate distributions over model parameters for system assessment. Our definition of subjective satisfaction is a direct consequence of combining quantified model uncertainty with SMC's satisfaction function.
%
%
Parametrized Bayesian model checking for DBNs \cite{langmead2009generalized} does deal with quantified model uncertainty. However, the author does not exploit the posterior for bounding or estimating errors. The algorithm terminates when the posterior variance ``is less than some user-specified threshold''. This approach does not yield statistical error estimates or bounds. We argue that in the context of software engineering, quantifiable error guarantees or estimates play a key role for system assessment.
%
%
A quite different approach to quantitative system assessment under model uncertainty is formal verification with confidence intervals (FACT) \cite{calinescu2016formal}. It is based on (exhaustive) probabilistic model checking, and therefore allows to perform more thorough analysis than BV, which is approximate and (temporally) bounded. However, for the same reason, FACT suffers from the state space explosion. FACT models uncertainty in terms of frequentist confidence intervals, in contrast to BV's Bayesian modeling approach.

\section{Conclusion}
\label{sec:conclusion}

We have presented a Bayesian approach to statistical model checking under model uncertainty. We introduced the notion of subjective satisfaction as a result of combining recently introduced satisfaction functions with model uncertainty. We also presented Bayesian Verification (BV), an approximate Monte Carlo style algorithm for assessing subjective system satisfaction based on a simulation. BV allows for user-specified confidence bounds, and thus enables to statistically bound verification errors. We empirically evaluated BV on a toy example with positive results.

There are some limitations to the BV algorithm. When $p_\mathrm{sat}$ is close to $p_\mathrm{req}$, BV may take a many iterations to establish the required confidence. Note that this property is independent from the absolute value of $p_\mathrm{sat}$. Similar to Bayesian model checking based on a fixed model, BV scales well with required satisfaction probabilities close to one (see e.g. \cite{zuliani2010bayesian}). BV's obtained error bounds are statistical: They do not provide a hard upper bound. I.e. this bound may be surpassed temporarily when operating BV (e.g. an error may occur even when running BV only once, yielding an error rate of one). Also, while we could empirically observe that the error bound was not severely violated for our toy problem, there may be an intimate connection to the choice of prior for $P(\theta)$. To study the connection of prior and error bound would probably yield interesting directions for further research. Another limitation of BV is its boundedness in terms of search depth. To this end, it would be interesting to increase the quality of satisfaction estimates, for example by adding global, previously trained satisfaction estimators to BV.


\section*{Acknowledgment}

The authors would like to thank Martin Wirsing and Matthias H\"olzl for many inspiring discussions that led us into the direction of research presented in this paper.



\bibliographystyle{IEEEtran}
\bibliography{IEEEabrv,references}
%
%
%

\end{document}